\documentclass[conference]{IEEEtran}
\IEEEoverridecommandlockouts
\usepackage{color}
\usepackage{cite}
\usepackage[numbers]{natbib}
\usepackage{amsmath,amssymb,amsfonts}
\usepackage{algorithmic}
\usepackage{graphicx}

\begin{document}

\title{Intrusion Detection Systems for Networked Unmanned Aerial Vehicles: A Survey\\
{\footnotesize \textsuperscript{}}
\thanks{}
}

\author{\IEEEauthorblockN{Gaurav Choudhary, Vishal Sharma, \\Ilsun You, Kangbin Yim}
\IEEEauthorblockA{\textit{Department of Information Security Engineering} \\
\textit{Soonchunhyang University, ROK}\\
Email: \{gauravchoudhary7777, ilsunu\}@gmail.com,\\vishal\_sharma2012@hotmail.com, yim@sch.ac.kr }

\and
\IEEEauthorblockN{Ing-Ray Chen}
\IEEEauthorblockA{\textit{Department of Computer Science} \\
\textit{Virginia Tech}\\
\textit{VA, USA} \\
Email: irchen@vt.edu}
\and
\IEEEauthorblockN{Jin-Hee Cho}
\IEEEauthorblockA{\textit{U.S. Army Research Laboratory} \\
\textit{MD, USA} \\
Email: jin-hee.cho.civ@mail.mil}
}

\maketitle

\begin{abstract}
Unmanned Aerial Vehicles (UAV)-based civilian or military applications become more critical to serving civilian and/or military missions. The significantly increased attention on UAV applications also has led to security concerns particularly in the context of networked UAVs. Networked UAVs are vulnerable to malicious attacks over open-air radio space and accordingly intrusion detection systems (IDSs) have been naturally derived to deal with the vulnerabilities and/or attacks. In this paper, we briefly survey the state-of-the-art IDS mechanisms that deal with vulnerabilities and attacks under networked UAV environments. In particular, we classify the existing IDS mechanisms according to information gathering sources, deployment strategies, detection methods, detection states, IDS acknowledgment, and intrusion types. We conclude this paper with research challenges, insights, and future research directions to propose a networked UAV-IDS system which meets required standards of effectiveness and efficiency in terms of the goals of both security and performance.
\end{abstract}


\section{Introduction} \label{sec:introduction}

The proliferation of unmanned aerial vehicles (UAVs) and their diverse applications in many different domains have been realized due to their merit of dynamic reconfigurability, fast response, and ease of deployment. In particular, the applications of networked UAVs have attracted major industry players such as Google, Facebook, Boeing, and Amazon. In addition, their applications in serving military and civilian missions have been explored in diverse domains to provide public safety, surveillance, medical services, and/or military mission support~\cite{Pajares15}. In Table \ref{table:UAV applications}, we discuss the key application domains where UAVs can be applied to assist given missions in different domain context. 

The key merit of UAVs is known as its high reconfigurability and mobility. However, its mobility also exposes an issue of controllability towards the aerial vehicles and causes link distortion in UAV networking. Despite these concerns, UAV-assisted networks have been recognized for the benefit of easy deployment of wireless connectivity that does not require any physical infrastructure~\cite{Sharma16, Sharma16a}. 

\begin{table*}
\centering
\caption{Domains and applications of UAVs} \label{table:UAV applications}
\begin{tabular}{p{4cm}p{6cm}p{7cm}}
\hline
\multicolumn{1}{c}{Domain} & \multicolumn{1}{c}{Key example applications} & \multicolumn{1}{c}{Achieved roles by UAVs} \\
\hline
Law enforcement surveillance & Search and rescue  & Equipped with camera \\
Public safety communications & Voice communications in case of disaster & Aerial base stations  \\
Environmental applications & Climate change & Equipped with sensors \\
Logistics &  Goods shipping / delivery in urban areas  & Drone as a transportation medium  \\
Military applications & Searches for lost or injured soldiers & Armed with live video remote communications to ground troops, essential gear, or weapons \\
Medical field applications & Delivering aid packages, medicines, vaccines, blood and other medical supplies to remote areas & Drone as a transportation medium \\
Video and photography & Events (e.g., social gatherings, sports games, or competitions) & Equipped with camera  \\
Agriculture & Crop monitoring and soil and field analysis & Equipped with sensors \\
\hline
\end{tabular}
\end{table*}

UAVs provide high benefits to assist the goals of many different applications, as summarized in Table \ref{table:UAV applications}. However, they also introduce the following challenges~\cite{Sharma17}: (1) the architectural design of drone communication lacks a standard or unification; (2) UAV-assisted communication networks suffer from an issue of dedicated spectrum sharing; (3) UAV deployment and path planning should be considered during spectrum allocations due to its potential impact on energy efficiency; and (4) UAV communications introduce additional overhead to architectural design, deployment, and consistency with large and reliable networks along with their security. In this work, we particularly focus on security challenges.

This paper provides the following {\bf key contributions}:
\begin{itemize}
\item We survey the key state-of-the-art UAV-IDS approaches and associated taxonomies which can provide a good overview to answer what a UAV-IDS system is, what the key components need to be considered in the UAV-IDS, and what the key security concerns should be considered, associated with the key components of the UAV-IDS. 
\item We discuss the main research challenges and hurdles to build a cyber-physical hardened UAV-IDS system under highly resource-constrained, hostile, dynamic, and distributed environments reflecting the key characteristics of military tactical characteristics.
\item We suggest future research directions to move towards based on the discussed research challenges and learned lessons / insights.
\end{itemize}

The remainder of the paper is organized as follows. Section \ref{sec:uav_ids_background} provides the background and goal of UAV-IDS. Section \ref{sec:texonomies} discusses the taxonomies used in the structure and classification of UAV-IDS. Section \ref{sec:evaluation} discusses the evaluation techniques of the state-of-the-art UAV-IDS approaches. Section \ref{sec:research_challenges} discusses research challenges derived from the inherent characteristics of UAV-IDS environments. Section \ref{sec: conclusion} concludes the paper and suggests future work directions.

\section{Overall Description of UAV-IDS Environments} \label{sec:uav_ids_background}

An unmanned aerial vehicle based intrusion detection system (UAV-IDS) is developed to detect anomaly behaviour or illegal activities in a network by automatically analyzing the behaviors or activities based on given hypothesis and/or policies, which are governed by the security rules of the given network~\cite{Biermann2001}. The UAV-IDS monitors system configuration, data files, and/or network transmission to check whether there exists an attack. Hence, the UAV-IDS is to mitigate the effect of the attacks aiming to prevent any covert / overt operations from exposed vulnerabilities of the system. In addition, UAV-IDSs aim to detect the misuse of UAVs. Misuse can be defined as any undesirable activity which can cause any harmful effect in terms of either performance or security to an entire swarm of the UAVs. Attacks explore the vulnerabilities of UAV systems, where the vulnerabilities can be the result of misconfigurations of UAV networks, an implementation fault, flawed designs and/or protocols~\cite{Debar99}.

\begin{figure}[ht!]
\centering
\includegraphics[width=250px]{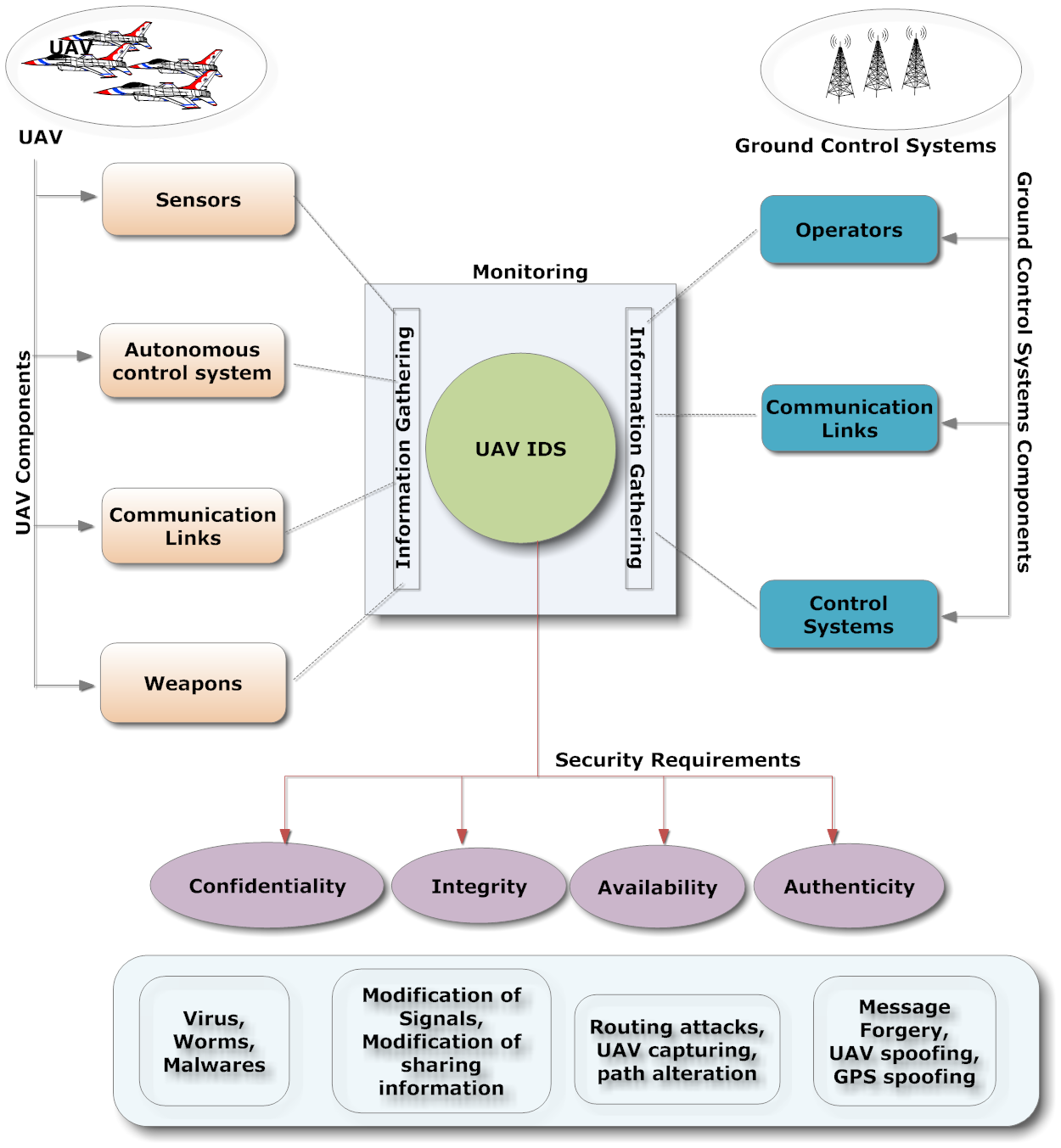}
\caption{An operational overview of UAV-IDS.}
\label{operational_overivew}
\end{figure}

Fig.~\ref{operational_overivew} shows an example scenario for an UAV-IDS. UAV-IDSs monitor signals, command traffic, control instructions, working behavior, energy consumption, and/or operations of UAV components. In addition, it analyzes the data flow and gather information from different components of UAVs during their operations as a network node. The UAV-IDSs are capable of enhancing reliability and/or security of UAV communications in an efficient and effective manner.

A UAV-IDS can be placed on a UAV or a ground control system and maintains the security and reliability of the UAV and the ground control system. The placement of a UAV-IDS can be determined based on the level of required security, such as required security levels in terms of confidentiality, integrity, availability, and authorization. In addition, a UAV-IDS is responsible for ensuring guarding the UAV system against unlawful activities or attacks. The incorporated security policies for UAVs provide low-complex rules to detect anomalies or potential threats. These policies can be designed through different approaches or algorithms based on the requirements of the UAV system. Most existing IDSs for UAVs use behaviour-based detection mechanisms~\cite{Shen2007}.

\section{Taxonomies of UAV-IDS System Components} \label{sec:texonomies}
We summarize the key component taxonomies of the UAV-IDS system in Fig. \ref{Taxonomy} which discusses its key components, including information gathering sources, deployment strategies, detection methods, detection states, acknowledgment, and intrusion types. We give the detail of each component as below.

\subsection{Information Gathering Sources}
A UAV is embedded with a cyber-physical system consisting of sensors and/or actuators. Sensors provide data (or information) to an actuator that can control the UAV. The collected data are used for analysis to make mission-related critical decisions. The information gathering sources can be classified as follows: \cite{Lauf2010} 
\begin{itemize}
\item {\bf Sensors}: Sensors collect information in terms of signals and/or behavior through sensors like inertial sensors, location sensors, and/or threat sensors. The sensor may be implicitly tied with a UAV or explicitly tied to a specific task object, like weather capturing sensors. The information retrieval by a malicious node from any of onboard sensors in a critical situation can impact the performance of a UAV in a networked scenario.
\item {\bf Communications links}: Communication links support transmissions directly to UAVs in mission areas and/or allow simultaneous sharing of information among multiple UAVs and the ground system. They also secure data transfers by monitoring the traffic between a source and a destination.
\item {\bf Ground control system (GCS)}: The GCS has a significant component in UAVs and is charged of conducting intelligent surveillance and reconnaissance based on data generated by the unmanned aircraft's payload.
\item {\bf UAV components}: The components within a UAV include a power supply unit, antennas, transceiver units, navigation systems, and an inbuilt UAV control system. All inter- and intra-communications take place through these components, in which the information is exchanged among these components for an effective control and maneuvering of UAVs. This information should be examined for security purposes and timely patches should be available upon the detection of potential threats.
\item {\bf Deployment strategies}: The deployment of an IDS in UAVs is critical because effective optimization is required for balancing the trade-off between IDS operations and UAV transmissions. The system should be maintained to enhance performance of the UAV with their effective operations and environmental conditions along with controllable activities of the deployed IDS. The IDS can be deployed based on two methods: 
\begin{enumerate}
\item {\bf Ground-coordinated or network-initiated basis}: In the ground coordinated IDS, all the gathered information is analyzed on the ground station and appropriate decisions are made on the basis of analyzed data; and 
\item {\bf Autonomous or host basis}: With an autonomous deployment of IDSs, UAVs acting as hosts to deploy IDSs should conduct data analysis and control other UAVs, along with coordinating between these two. In this deployment type, the IDS is placed within the system control of UAVs in the form of hardware or software.
\end{enumerate}
\end{itemize}

\subsection{Intrusion Detection Systems (IDSs)}
The key mechanism of IDSs can be classified as follows:
\begin{itemize}
\item {\bf Specification-based~\cite{Tseng2003}}: A UAV-IDS is incorporated with respective rules specified based on the expected behaviors of UAVs. These specified rules are applied to monitor successful executions of the UAV system.
\item {\bf Signature-based~\cite{Vaidya2001}}: This method aims to detect known attacks based on pre-defined, known signatures. Upon detecting anomaly activities, a detection operation is triggered to identify a matched signature to ensure the detection of an intrusion.
\item {\bf Anomaly-based~\cite{Patcha2007}}: Anomaly behavior is detected based on a failure or an illegal activity observed in a system. With the goal of detecting known and/or unknown attacks, this method uses learning or a filtering mechanism, which can significantly enhance the detection of unknown attacks in the absence of pre-defined signatures of the unknown attacks.
\item {\bf Hybrid-based~\cite{Aydin2009}}: This method is a hybrid approach by integrating two or more detection methods, such as specification plus anomaly, in order to provide a strong detection policy that can catch known and/or unknown attacks. 
\end{itemize}

\begin{figure}[ht!]
\centering
\includegraphics[width=250px]{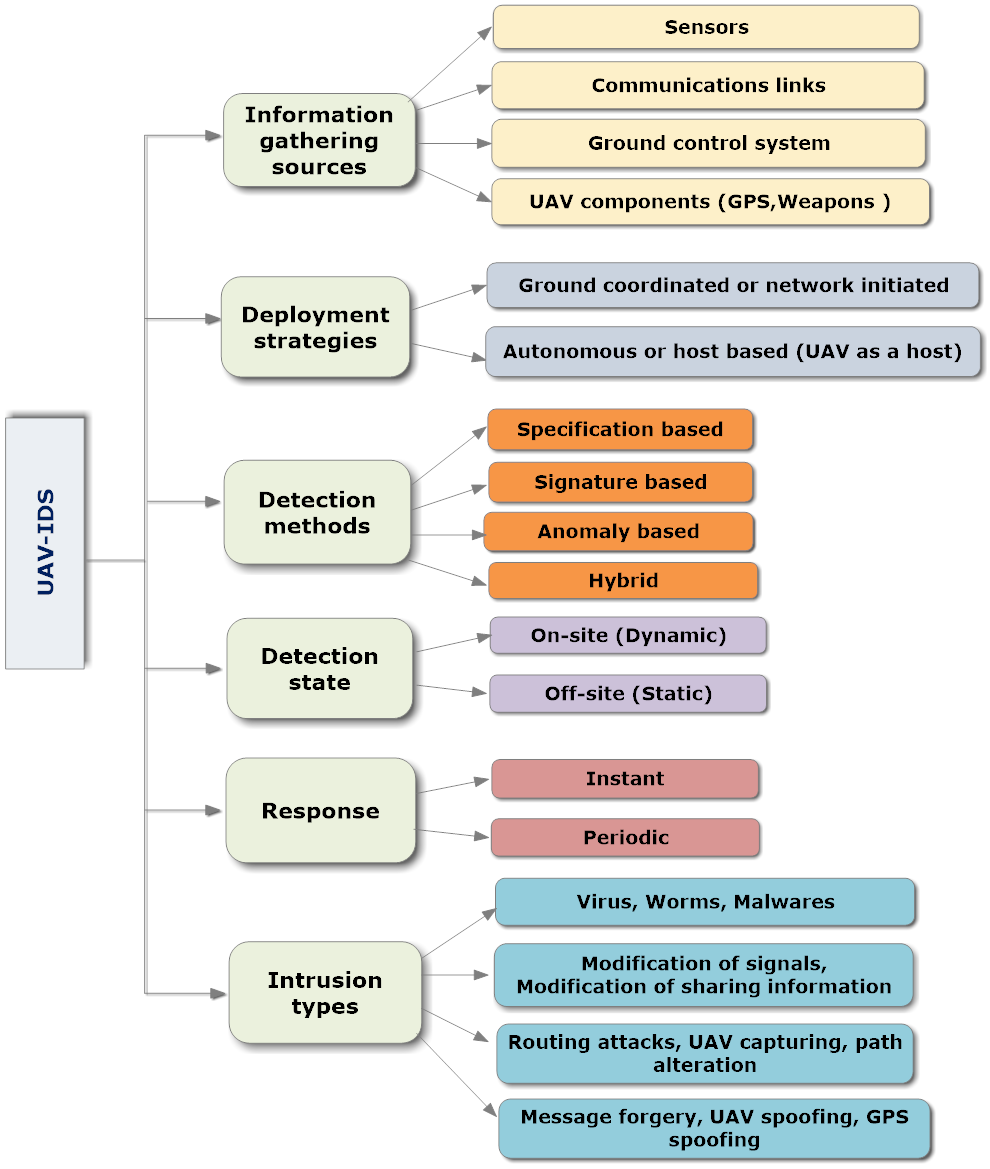}
\caption{Taxonomy of UAV-IDSs.}
\label{Taxonomy}
\end{figure}

\begin{table*}[!ht]
\centering
\caption{Survey on the State-of-The-Art on Existing UAV IDS Approaches.}
\label{my-label}
\vspace{-2mm}
\begin{tabular}{p{3cm}p{6cm}p{7cm}}
\hline
\multicolumn{1}{c}{Reference}   & \multicolumn{1}{c}{Proposed scheme}   & \multicolumn{1}{c}{Key method(s) / metrics}                                                                                                       \\ \hline 
\citet{Blasch2011}  & War planning situation awareness tool & ROC for visualization                                                                                                                                        \\
\citet{Lauf2010}  & Hybrid IDS      & Maxima and cross-correlation detection                                                                                                                      \\
\citet{Shen2007}  & Markov game theoretic approach   & Deployment of IDS, configuration of email-filtering, firewall settings, and shutdown or reset policies for servers. 
\\
\citet{Muniraj2017}     & Framework for detection of cyber-physical attacks on the sensors  &  Anomaly-based detectors based on knowledge of the physical system and statistical analysis.                \\
\citet{Sedjelmaci2017}  & Hierarchical IDS   & Threat classification \& behavior monitoring                                                                                                                    \\
\citet{Chen2014}  & Behavior rule-based evaluations    & Minimizing false positives \& false negatives                                                                                                                      \\
\citet{Kwon2015}  & Safety analysis under stealthy cyber attacks  & Real-time safety assessment                                                                                                                                       \\
\citet{Lauf2010a} & Distributed resource system    & Intrusion-tolerance strategy                                                                                                                         \\
\citet{Shen2008}  & Game theoretic approach    & Based on the three levels: object, situation and threat.  \\
\citet{sedjelmaci2017intrusion}  & Security game framework  & Optimal setting identification based on intrusion detection rate with minimum overhead using Bayesian game                                   \\
\citet{Trafton2006}  & Network service suite & Framework for information assurance of UAVs                                                                                                                      \\ \hline
\end{tabular}
\vspace{-5mm}
\end{table*}

\subsection{Detection States}
Based on the source of information, we can categorize two major detection states: 
\begin{itemize}
\item {\bf On-site (or dynamic)}: The detection state is evaluated based on data collected and monitored from real-time operations; the detection analysis and decisions are made at on-site UAVs.
\item {\bf Off-site (or static)}: The detection state is evaluated based on data collected by an IDS from all information sources; the detection is made based on the analysis of all collected data received in the IDS.
\end{itemize}

\subsection{IDS Acknowledgment}
Based on the result of data analysis, a UAV-IDS makes decision on whether there exists an attack via an IDS acknowledgment. This IDS acknowledgment has two forms: 
\begin{itemize}
\item {\bf Instant acknowledgement}: In this acknowledgment, an IDS monitoring is performed at real-time and decisions or alarms are generated in the form of instant acknowledgments.
\item {\bf Periodic acknowledgement}: In this acknowledgement, an IDS continuously gathers the data but the decisions are based on the periodical analysis of received data.
\end{itemize}

\subsection{Intrusion Types}
A UAV-IDS should be able to detect the following intrusion types:
\begin{itemize}
\item Virus, worms, and/or malware;
\item Modification of signals, modification of sharing information;
\item Routing attacks, UAV capturing, and/or path alteration; and
\item Message forgery, UAV spoofing, and/or GPS spoofing.
\end{itemize}

\section{Survey of Existing UAV-IDS Approaches} \label{sec:evaluation}

UAV networks are highly sensitive over which critical information will be exchanged between UAVs and the ground station. Table~\ref{Taxonomy} summarizes the state-of-the-art UAV-IDS approaches, aiming to enhance security and performance with the end goal to build a cyber-physical hardened system being protected against inside and outside attackers on networked UAVs. Below we survey these existing approaches using our proposed taxonomies discussed in Section \ref{sec:texonomies}.

\citet{Blasch2011} proposed a war planning situation awareness tool by leveraging the Receiver Operating Characteristic (ROC) plots to visualize the effectiveness of their classifications. The effective classification is developed based on matrices where situation assessment is used to derive relations between a given classification and a location. \citet{Lauf2010} developed a decentralized anomaly-based detection technique, which uses maxima and cross-correlation detection methods. The Maxima Detection System (MDS) allows the characterization of either one or zero suspicious nodes. Cross-correlation detection methods are capable of detecting multiple intrusions. However, this work does not capture the quality of the IDS based on detection errors including false positives and false negatives.

\citet{Shen2007} took a game theoretic approach by considering three levels of states: object, situation, and threat. This approach projects attack activities while focusing on the states of the network. \citet{Shen2008} further developed a cooperative surveillance strategy to improve the performance through adaptive Markov game based on the cooperative jamming strategies. These are performed on the basis of four defensive parameters, including IDS deployment, configurations of email-filtering, firewall settings, and shut down or reset policies for servers.

\citet{Muniraj2017} focused on the attacks over small UAVs by identifying malicious activities over their sensors. In the proposed framework that detects cyber-physical attacks, sensors are designed based on the knowledge of physical system and statistical analysis techniques. However, the proposed scheme was not capable of detecting combination of piece-wise constant attacks of smaller magnitude. \citet{Sedjelmaci2017} proposed an hierarchical IDS and intrusion response mechanism by classifying threats and monitoring UAV behavior to detect malicious activities.

\citet{Chen2014} proposed a specification-based detection technique to guard a UAV system against cyber-attacks. This work used a behavior rule-based UAV-IDS, in which the behavior rules are constructed based on defined attack models, considering reckless, random, and opportunistic attacks. This work minimized detection errors (i.e., false positives and false negatives) based on the critical tradeoff between security and performance of UAVs. \citet{Kwon2015} developed a real-time safety assessment algorithm based on reachability analysis to deal with cyber attacks.

\citet{Lauf2010a} developed a distributed sensing mechanism to build a fault-tolerant resource management system. The proposed mechanisms uses a service discovery protocol (SDP) where SDP flooding can introduce a burst of communications leading to traffic congestion or bottleneck issues. \citet{Sedjelmaci2016a} proposed a threat estimation model based on estimated beliefs towards whether a threat exists in the system. In addition, this work incorporated specific detection policies to maintain data integrity and network availability. \citet{sedjelmaci2017intrusion} took one step further to propose a robust UAV assisted network against lethal attackers, namely a Security Game Framework (SGF), which is formulated based on Bayesian game among the suspected nodes. This approach formulates two attack-defense problems between IDS and the attacker, and between intrusion ejection system and the suspected nodes.

\citet{Trafton2006} proposed the so called Joint Airborne Network Services Suite, which aims to integrate an airborne military network by allowing the implementation of various possible hardware and software solutions. In this work, an IDS is considered as an integral part of their assurance strategy.

\section{Research Challenges} \label{sec:research_challenges}

UAVs are operated remotely while receiving control and command messages from ground stations. These command and control messages are transmitted over different channels and variable transmission rate. Security vulnerabilities can be exploited to compromise confidentiality, integrity, availability, and authorization of networked UAVs~\cite{Javaid12,he2017communication}. Message security and control signal protections are achieved by cryptographic mechanisms. However, security issues, like unauthorized access, malicious control, illegal connection, or other malicious attacks, require strategic solutions without compromising performance. Identifying and mitigating threats in UAV networks efficiently and effectively is a first step to secure UAV networks~\cite{Sharma17a, Sharma17c}.

The significant increase of threats and/or attacks in UAV networks brought our attention on the issue of the deployment of IDS which will play a key role to achieve the effectiveness and efficiency of the UAV-IDS~\cite{Teodoro2009}. In a UAV environment, an IDS is being operated based on specific rules and/or policies to determine whether an observed activity is malicious or not. The results of the IDSs can be used to develop strategies to mitigate the identified risks. However, the design and development meeting these two requirements (i.e., effectiveness and efficiency of the developed IDS) is not a trivial goal because it often requires a time-consuming, high-overhead process which can often exceed the benefit of introducing high security in practice. 

To achieve the UAV-IDS system that meets required levels of effectiveness (i.e., minimizing detection errors with minimum service interruptions) and efficiency (i.e., reducing computational and communication overhead), we identify the following challenges on the table to pave a way to build a cyber-physical hardened UAV-IDS system:
\begin{itemize}
\item {\bf Detection latency}: The detection latency can be used as a measure of agility of an IDS. However, there is a critical tradeoff in that triggering the IDS more often leads to incurring more communication/computation overhead, which naturally results in low efficiency, and vice-versa. Hence, we need to make a good balance to achieve both efficiency and effectiveness in order to build an affordable, secure UAV-IDS systems in practice.
\item {\bf IDS computational cost}: The computational cost associated with IDSs is closely related to how much we want to achieve the accuracy of the IDS and security vulnerabilities we allow in a given system. Again, this issue is not trivial because more cost not only incurs high overhead, but brings more benefit in enhancing security.
\item {\bf Implementation overhead}: The high implementation overhead of IDSs causes power consumption and degrades the performance of UAVs resulting in a network shutdown.
\item {\bf Threat \& behavior modeling}: IDS detection techniques incorporate behaviors of UAVs. The rules are designed by reflecting the UAVs' behavior and/or possible threats. However, accurate observations of threat/attack behaviors and accordingly their correct modeling is not a trivial task although achieving it can provide an enormous benefit to expedite the development of better defense strategies.
\item {\bf Effective threat assessment}: An effective threat assessment is critical to mitigating vulnerabilities and risks associated with threats occurred. In particular, developing effective threat assessment policies is the key to enhance both security and performance of UAV-IDS systems.
\item {\bf Maximum network throughput with minimum cost}: This is a typical tradeoff issue any network can face as these two goals are conflicting to each other. However, based on dynamic monitoring to capture an accurate system state, both goals that are dynamically set can be achievable.
\item {\bf Lightweight IDS with minimum resource consumption}: As UAVs are battery-operated and resource-constrained, the development of lightweight IDS mechanisms is highly challenging but a must to achieve in networked UAVs.
\item {\bf Effective monitoring and attack response}: The response against an attack is naturally related to how quickly the attack is detected by a given IDS. This implies that the effectiveness of the IDS is closely related to how quickly the system can respond to the detected attack. This is indeed the issue of the agility of a system which should take appropriate actions in order to minimize damages or vulnerabilities caused by the intrusion which exploits system vulnerabilities. The contested nature of UAV environments, characterized by resource-constraints, high hostility, high dynamics, and distributed nature, also adds more challenges to achieve this goal.
\end{itemize}

\section{Conclusion \& Future Research Directions} \label{sec: conclusion}
In this work, we provided a brief overview of the state-of-the-art UAV-IDS mechanisms. In addition, we discussed related design challenging issues to develop effective and efficient UAV-IDS mechanisms, considering high resource-constraints, high hostility characterized by sophisticated attack/threat behaviors, and distributed nature causing high security vulnerabilities. We also defined the taxonomies to describe the key components of UAV-IDS systems based on the state-of-the-art existing works. Lastly, we discussed key research challenges that should be considered for future research plans, aiming to build an affordable, secure cyber-physical UAV-IDS system.

As future work directions, we plan to conduct the following:
\begin{itemize}
\item {\bf Define an attack model that captures key attack behaviors targeting for UAV-IDS systems}. We will derive an attack graph and build a set of corresponding countermeasures to deal with those attacks.
\item {\bf Develop a behavior rule specification-based UAV-IDS} that uses minimum memory while maximizing detection accuracy by checking the anomaly of an observed behavior. Formal verification and Bayesian estimation based ground truth check for anomaly behaviors can be used to validate the developed set of specification rules.
\item {\bf Measure the effectiveness and efficiency of the developed lightweight UAV-IDS} using the metrics of agility or resilience.
\end{itemize}

\section*{Acknowledgment}
This work was supported by the Institute for Information \& communications Technology Promotion (IITP) grant funded by the Korea government (MSIP) (2015-0-00508, Development of Operating System Security Core Technology for the Smart Lightweight IoT Devices). This work is also supported in part by the U.S. AFOSR under grant number FA2386-17-1-4076. Dr. Ilsun You is the corresponding author.

\bibliographystyle{IEEETranSN}
{
\small
\bibliography{uav-ids-bib}
}

\end{document}